\documentclass[aps,prl,twocolumn,groupedaddress,amsmath,superscriptaddress]{revtex4-1}

\usepackage{amsfonts}
\usepackage{amssymb}
\usepackage{amsmath}
\usepackage[dvips]{graphicx}
\usepackage{color}

\usepackage{amsmath}
\usepackage{graphicx}
\usepackage{amsfonts}
\usepackage{amssymb}
\usepackage{hyperref}
\usepackage{color}
\usepackage{mathrsfs}
\usepackage{isomath}
\usepackage{amsmath}
\usepackage{amsthm}
\usepackage{epstopdf}
\usepackage{txfonts}
\usepackage{verbatim}

\begin{document}

\title{\hspace*{-.6cm}$\mbox{Quantum-enhanced stochastic phase estimation with SU(1,1) interferometer}$}
\author{Kaimin Zheng}
\address{National Laboratory of Solid State Microstructures, Key Laboratory of Intelligent Optical Sensing and Manipulation, College of Engineering and Applied Sciences, and Collaborative Innovation Center of Advanced Microstructures, Nanjing University, Nanjing 210093, China}
\author{Minghao Mi}
\address{National Laboratory of Solid State Microstructures, Key Laboratory of Intelligent Optical Sensing and Manipulation, College of Engineering and Applied Sciences, and Collaborative Innovation Center of Advanced Microstructures, Nanjing University, Nanjing 210093, China}
\author{Ben Wang}
\address{National Laboratory of Solid State Microstructures, Key Laboratory of Intelligent Optical Sensing and Manipulation, College of Engineering and Applied Sciences, and Collaborative Innovation Center of Advanced Microstructures, Nanjing University, Nanjing 210093, China}
\author{Liyun Hu}
\address{Center for Quantum Science and Technology, Jiangxi Normal University, Nanchang 330022, People¡¯s Republic of China}
\author{Shengshuai Liu}
\address{State Key Laboratory of Precision Spectroscopy, School of Physics and Electronic Science, East China Normal University, Shanghai 200062, China }
\author{Yanbo Lou}
\address{State Key Laboratory of Precision Spectroscopy, School of Physics and Electronic Science, East China Normal University, Shanghai 200062, China }
\author{Jietai Jing}
\email{jtjing@phy.ecnu.edu.cn}
\address{State Key Laboratory of Precision Spectroscopy, School of Physics and Electronic Science, East China Normal University, Shanghai 200062, China }
\address{Collaborative Innovation Center of Extreme Optics, Shanxi University, Taiyuan, Shanxi 030006, People¡¯s Republic of Chin }
\author{Lijian Zhang}
\email{lijian.zhang@nju.edu.cn}
\address{National Laboratory of Solid State Microstructures, Key Laboratory of Intelligent Optical Sensing and Manipulation, College of Engineering and Applied Sciences, and Collaborative Innovation Center of Advanced Microstructures, Nanjing University, Nanjing 210093, China}

\begin{abstract}
The quantum stochastic phase estimation has many applications in the precise measurement of various physical parameters. Similar to the estimation of a constant phase, there is a standard quantum limit for stochastic phase estimation, which can be obtained with the Mach-Zehnder interferometer and coherent input state. Recently, it has been shown that the stochastic standard quantum limit can be surpassed with non-classical resources such as the squeezed light. However, practical methods to achieve the quantum enhancement in the stochastic phase estimation remains largely unexplored. Here we propose a method utilizing the SU(1,1) interferometer and coherent input states to estimate a stochastic optical phase. As an example, we investigate the Ornstein-Uhlenback stochastic phase. We analyze the performance of this method for three key estimation problems: prediction, tracking and smoothing. The results show significant reduction of the mean square error compared with the Mach-Zehnder interferometer under the same photon number flux inside the interferometers. In particular, we show that the method with the SU(1,1) interferometer can achieve the fundamental quantum scaling, the stochastic Heisenberg scaling, and surpass the precision of the canonical measurement.

\end{abstract}

\maketitle
\textit{Introduction}$-$The quantum optical phase estimation is a critical task in many applications, such as quantum imaging \cite{Taylor2013,Ono2013,Brida2010}, quantum sensing \cite{Pirandola2018,Degen2017,Bonato2015},
gravitational wave detection \cite{Adhikari2014,Ma2017}. To date most of works focus on the estimation of a constant phase $\varphi$, in which Mach-Zehnder interferometer (MZI) is the most commonly used device \cite{Pezze2008,Holland1993}. The precision of estimation is limited by the shot noise when the classic resources are used. This limit is often called the standard quantum limit (SQL): $ \triangle\varphi\propto1/\sqrt{N}$, where $N$ is the average number of photons in the probe state \cite{Caves1981,Giovannetti2011}. Many efforts have been taken to improve the precision. Most of them focus on utilizing nonclassical states to reduce the
quantum noise, such as the squeezed states and entanglement states \cite{Thomas-Peter2011,Xiao1987,Rosen2012}. It has been shown that the maximally entangled number
state (N00N) is the optimum probe state to reach the Heisenberg limit (HL): $\triangle\varphi\propto1/{N}$ \cite{Boto2000,Israel2014}. Moreover, for the constant phase estimation, the variance of estimation ($1/\sqrt{vN}$ or $1/\sqrt{v}N$) will decrease indefinitely as the number of measurement $v$ increases.

However, it is not enough to just estimate the constant phase because many signals of interest in real world are time-varying and stochastic \cite{Berry2002,Iwasawa2013,Tsang2011,Miao2017,Jimenez-Martinez2018}. Thus how to estimate such a time-varying phase with high precision is of practical importance. Assuming $\varphi\left( t\right) $ is the phase to be estimated. It can be treated as a constant in $t_i\leq t <t_i+dt$ if the $dt$ is small enough, i.e., $\varphi\left( t\right) =\varphi_i$, so the phase can be discrete as $\left(\varphi_0,\varphi_1,...,\varphi_{i},...,\varphi_{n-1},\varphi_n\right)$.
There are a set of observations $\left(r_0,r_1,...,r_i,...,r_{n-1},r_{n}\right) $ used to estimate the $\{\varphi_i\}$, where $r_i$ is the observation at time $i$. Compared to the estimation of constant phase, $ \left\langle \left(\varphi_i -\varphi_{i\pm s} \right) ^{2}\right\rangle$ between the phases at two different times ($\varphi_i,\varphi_{i \pm s}$) increases as the $\left|s\right|$ increases. Therefore, the correlation between $r_{i\pm s}$ and $\varphi_i$ decreases as $s$ increases and the number of observations can be used to improve precision are limited. So there is a limited precision for the stochastic phase estimation even for infinite measurement time.
It has been shown that the mean square error (MSE) in the estimation of a stationary Gaussian stochastic phase with a power-law spectrum $\kappa^{p-1}/\left(\omega ^{p}+\lambda ^{p}\right)$ using coherent states scales as $ \left(\kappa/\mathcal{N}\right)^{\left(p-1\right)/p}$, which is called stochastic SQL. Here, $ \mathcal{N}$ is the photon flux \cite{Berry2015, Berry2013,Dinani2017}. Similar to constant phase estimation, there is a stochastic Heisenberg scaling for the estimation of this stochastic phase which scales as  $ \left(\kappa/\mathcal{N}\right)^{2\left (p-1\right)/\left(p+1\right)} $ \cite{Berry2015,Berry2013,Dinani2017}. Previous works show that the stochastic SQL can be surpassed with non-classical resources such as the squeezed light with adaptive quantum smoothing technique or canonical phase measurement \cite{Tsang2009,Tsang2009a,Tsang2009b,Berry2006,Wheatley2010,Yonezawa2012}. However, practical methods to achieve the quantum enhancement in the stochastic phase estimation still remains largely unexplored.

In this paper, we propose a method to estimate the stochastic phase using a new measurement device, the SU(1,1) interferometer. Such device, also known as the nonlinear interferometer (NLI), has been proposed recently and experimentally demonstrated with a signal to noise ratio (SNR) surpass that of the MZI \cite{Jing2011,Ou2012,Hudelist2014,Anderson2017,Manceau2017,Lemieux2016}. By combining the NLI with adaptive feedback technique, we show that MSE of the estimation is reduced over a range of parametric amplifier gain $G$ compared to the that of MZI. For a fixed photon flux, there is an optimal $G$ minimizes the MSE and maximizes the precision. In particular, with the optimal $G$ the precision of our scheme surpasses the stochastic SQL and achieves the stochastic Heisenberg scaling asymptotically.

\begin{figure}[h!]
\begin{center}
\includegraphics[width=80mm]{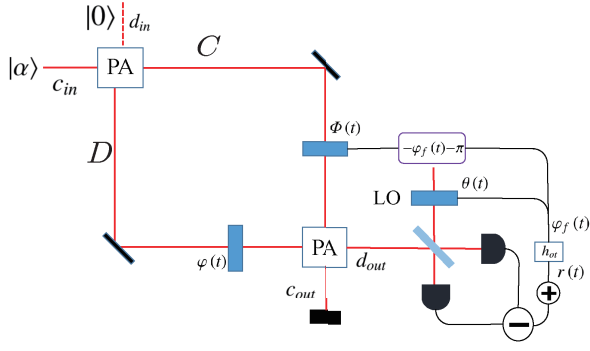}
\end{center}
\caption{The schematic diagram of enhanced stochastic phase estimation with SU(1,1) interferometer. This interferometer is consists of two parametric amplifiers (PA) and the input states are coherent state and vacuum
state. The $\varphi\left( t\right) $ is the stochastic phase to be estimated and the phase $\Phi (t)$ in the other arm was adaptively controlled. $r(t)$ is photocurrent which is equal to the homodyne measurement results after an added operation. The phase $\theta (t)$ of the local oscillator was adaptively controlled simultaneously and $h_{ot}$ is the optimum linear processor of phase tracking.}  \label{fig:1}
\end{figure}

\textit{Stochastic phase estimation scheme}$-$The schematic diagram of the estimation of a stochastic phase with NLI is shown in Fig .\ref{fig:1}, in which the
NLI contains two parametric amplifiers (PAs). Two input modes of the first PA are injected with a coherent state $\left\vert \alpha \right\rangle $ and
a vacuum state. The phase $\varphi\left( t\right)$ to be estimated is imposed to one arm of the interferometer. One of the output mode is measured with the homodyne measurement. The measured results after a displaced operation yield photocurrent $r(t)$. The phase $\Phi(t)$ in the other arm and the phase $\theta (t)$ of the local oscillator are adaptively controlled
based on $\varphi _{f}\left( t\right) $ which is estimated from $r(s)$ for all the region $s<t$.
In the NLI, the first PA plays the role of beam
splitting. If we define $\hat{c}_{in}, \hat{d}_{in}$ to be the
annihilation operators of the two inputs, and $\hat{C}, \hat{D}$ to be
the annihilation operators of the outputs, the relation of input-output of the PA can be written as $\hat{C}=G\hat{c}_{in}+g\hat{d}_{in}^{\dag }, \hat{D}=G\hat{d}_{in}+g\hat{c}_{in}^{\dag },$
where $G$ is the gain of the PA, and $G^{2}-g^{2}=1$ \cite{Ou2012}.
The second PA which has the same gain $G$ acts as the role of recombination, so the
complete input-output relation of the NLI is
\begin{align}
\hat{c}_{out}=&G\left( G\hat{c}_{in}+g\hat{d}_{in}^{\dag }\right)
e^{i\Phi \left( t\right) }+g\left( g\hat{c}_{in}+G\hat{d}_{in}^{\dag
}\right) e^{-i\varphi \left( t\right) },  \nonumber \\
\hat{d}_{out}=&g\left( G\hat{c}_{in}^{\dag }+g\hat{d}_{in}\right)
e^{-i\Phi \left( t\right) }+G\left( g\hat{c}_{in}^{\dag }+G\hat{d}%
_{in}\right) e^{i\varphi \left( t\right) }.
\end{align}
 When we perform a homodyne detection at the output mode $d_{out}$, and the homodyne detection result is added by $2Gg\left\vert \alpha \right\vert\varphi _{f}\left( t\right)$,
 the photocurrent can be approximately represented as \cite{supplementary}
\begin{equation}
r\left( t\right) \approx\frac{2Gg\left\vert \beta \right\vert}{\sqrt{G^2+g^2}} \varphi \left( t\right) +%
\sqrt{2G^{2}g^{2}\sigma _{f}^{2}+1}n\left( t\right),   \label{eq2}
\end{equation}
where we have adaptively controlled the feedback phase as $\Phi \left(
t\right) =-\varphi _{f}\left( t\right) -\pi ,$ and controlled the phase of
the local oscillator to be $\theta \left( t\right) =\varphi _{f}\left(
t\right)+\pi/2$, which aims to make the each measurement most sensitive and maximize the
phase information obtained. $n\left( t\right) $ is the normalized Gaussian white noise from the homodyne
measurement which satisfies $\left\langle n\left( t\right)n\left( s\right) \right\rangle =\delta \left( s-t\right) $. Moreover, the photocurrent has gaussian stationary statistic and $\sigma _{f}^{2}=\left\langle \left( \varphi \left( t\right) -\varphi_{f}\left( t\right) \right) ^{2}\right\rangle$ is stationary MSE and we have defined that the  photon flux in the interferometer is $\left\vert \beta \right\vert ^{2}=\left( G^{2}+g^{2}\right) \left\vert\alpha \right\vert ^{2}$.

 According to the time span of the observations which is used for the estimation, the time-varying phase estimation can be divided into three cases. Prediction: the future phase $\varphi_{i+m}$ is estimated with observations $\left(r_0,r_1,...,r_i\right) $. Tracking: the current and previous observations $\left(r_0,r_1,...,r_i\right) $ are used to estimate the current phase $\varphi_{i}$. Smoothing: the measurement results beyond the time $i$ are also used to estimate phase $\varphi_{i}$. To analyze these three estimation problems simultaneously, we introduce a general desired signal $ d\left( t\right) =\varphi \left( t+\varepsilon \right) $, which is estimated with photocurrent $r \left(s\leq t\right)$. Here, $\varepsilon$ can be any real number, and the three kinds of phase estimation can be defined according to the the value of $\varepsilon$. 
 Base on the measurement photocurrent $r(t)$, the desired signal can be estimated as
\begin{equation}
d_{f}\left( t\right) =\int_{-\infty }^{t}d\tau h_{o}\left( t,\tau \right)
r\left( \tau \right),   \label{3}
\end{equation}%
where $h_{o}\left( t,\tau \right)$ is the impulse response function which represents the output at time $t$ if the input at time $\tau$ is an impulse, and it can be marked as $h_{op}\left( t,\tau \right),h_{ot}\left( t,\tau \right),h_{os}\left( t,\tau \right)$ for prediction, tracking and smoothing, respectively. The optimum linear processor $h_{o}\left( t,\tau \right)$ for minimizing the MSE $%
\xi \left( t\right) =\left\langle \left[ d\left( t\right) -d_{f}\left(
t\right) \right] ^{2}\right\rangle $ satisfies \cite{Trees2013}
\begin{equation}
K_{dr}\left( t-\eta\right) =\int_{-\infty }^{t}h_{o}\left( t-\epsilon \right)
K_{r}\left( \epsilon -\eta \right) d\epsilon,   \label{eq4}
\end{equation}%
which is the Wiener-Hopf equation and $K_{dr}\left( t-\eta \right) =$ $%
\left\langle d\left( t\right) r\left( \eta \right) \right\rangle $, $%
K_{r}\left( \epsilon -\eta \right) =$\ $\left\langle r\left( \epsilon \right)
r\left( \eta \right) \right\rangle $. The correlation function only depends on the time difference due to the fact that received photocurrent and desired signal are jointly stationary and time-invariant. We can obtain the optimum linear impulse response function $h_{o}$ by solving the Wiener-Hopf equation, and the MSE of the phase estimation can be calculated at the same time.
\begin{figure}[tbp]
\begin{center}
\includegraphics[width=80mm]{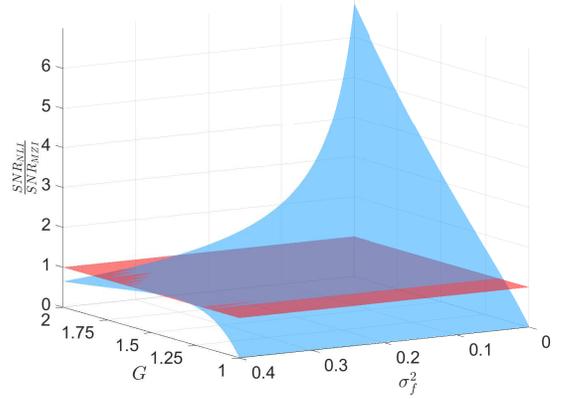}
\end{center}
\caption{The ratio of the two SNRs. Blue surface represent the ratio of the two SNRs and red surface represent the case of the SNR of two interferometers is equal.}
\end{figure}

In the MZI case, the splitting and recombining of light are accomplished by 50:50 beam spliters. The relation of input-output of the beam spliter is $\hat{C}=1/\sqrt{2}\left(\hat{c}_{in}+i\hat{d}_{in}\right)$ and $\hat{D}=1/\sqrt{2}\left(i\hat{c}_{in}+\hat{d}_{in}\right)$. Similar to NLI, we set the two feedback phases $\Phi \left(
t\right) =\varphi _{f}\left( t\right) $ and $\theta \left( t\right) =\varphi _{f}\left(
t\right)+\pi$ for the most sensitive estimation. The photocurrent can be calculated as
$r\left( t\right) \approx\left\vert \beta  \right\vert \varphi \left( t\right)+n\left( t\right)$ \cite{supplementary}.
Here, we should note that the two input modes of the first BS are injected with a coherent state $\left\vert \beta\right\rangle $ and
a vacuum state, which makes the photon number flux inside both interferometers the same. From the two photocurrents, we can derive the relation $\frac{SNR_{NLI}}{SNR_{MZI}}=\frac{4G^2\left( G^2-1\right)}{{\left( 2G^2-1\right)\left( 2G^2\left( G^2-1\right)\sigma _{f}^{2}+1\right)}}$,
where $SNR_{NLI}$ and $SNR_{MZI}$ are the SNRs of these two interferometers respectively. Fig.2 shows that the ratio of the two SNRs (Blue surface). As $G$ increases from 1, the SNR of the NLI increases with $G$, which agrees with the previous analysis of NLI \cite{Ou2012}. The SNR of the NLI surpasses that of the MZI only when $G$ is beyond certain threshold due to the asymmetric nature of the NLI. For finite $\sigma _{f}^{2}$, the further increase of $G$ will reduce the SNR. This result can be understood from Eq. (2): when $G$ is large, the signal term increases linearly with $G$ while the noise term increases quadratically with $G$. Since the MSE in the estimation of time-varying phase can not be arbitrarily small, we expect there is an optimal $G$ for stochastic phase estimation which is different from the case of measuring a constant phase.

\textit{Ornstein-Uhlenback stochastic phase estimation}$-$ As an example, we consider the situation that the time-varying phase $\varphi\left( t\right) $  to be estimated follows an Ornstein-Uhlenback stochastic process, which can be found in many practical physical process and defined by \cite{Wheatley2010}
\begin{equation}
\frac{d\varphi \left( t\right) }{dt}=-\lambda \varphi \left( t\right) +\sqrt{%
\kappa }\frac{dV\left( t\right) }{dt},  \label{1}
\end{equation}%
 where $\lambda ^{-1}$ is the correlation time of $\varphi \left( t\right) $, $%
dV\left( t\right) $ represents the Wiener process which satisfies $%
\left\langle dV\left( t\right) dV\left( s\right) \right\rangle =\delta
\left( s-t\right)dt $, and $\kappa $ is the magnitude of the
Wiener noise. The expectation value of $\varphi\left( t\right) $ is 0 and its statistics is stationary, which means the correlation between the phases at two different time only depends on their time difference. Moreover, the spectral density spectrum of $\varphi\left( t\right) $ is $S_{\varphi}\left( \omega
\right)={\kappa }/\left( {\omega ^{2}+\lambda ^{2}}\right)$. In this situation, Eq. (\ref{eq4}) can be solved with the Wiener technique, and the Fourier transform of the optimum linear response function $h_{o}$ is \cite{supplementary}
\begin{align}
H_{o}(\omega )&= \left\{
\begin{array}{lr}
\frac{\kappa\sqrt{P}e^{^{i\omega\varepsilon}}}{N\lambda\left(1+\sqrt{
1+\Lambda}\right)\left(\lambda\sqrt{1+\Lambda}+i\omega\right)}&,\varepsilon >0 \\%
\frac{\kappa\sqrt{P}e^{i\omega\varepsilon}}{N\left(\lambda^{2}\left(
1+\Lambda \right)+\omega^{2}\right)}\left[1-\frac{e^{\varepsilon\left(
\lambda\sqrt{1+\Lambda}-i\omega\right)}\left(\lambda+i\omega\right)}{
\lambda\left(1+\sqrt{1+\Lambda}\right)}\right]&,\varepsilon \leq 0
\end{array},
\right.   \label{eq6}
\end{align}
where $P=4G^{2}g^{2}\left\vert \alpha \right\vert ^{2}$, $\Lambda =%
{P\kappa }/{N\lambda ^{2}}$, $N=2G^{2}g^{2}\sigma _{f}^{2}+1$. Eq. (\ref{eq6}) shows that the optimum linear processor is a low-pass filter with a cut-off frequency of $\lambda \sqrt{1+\Lambda }$ in the case of $\varepsilon =0$.
When this optimum linear processor is used to estimate the phase, the minimum MSE of the estimation is \cite{supplementary}
\begin{equation}
\xi
=\left\langle \left[ d\left( t\right) -d_{f}\left( t\right) \right]
^{2}\right\rangle=K_{d}\left( 0\right) -\int_{0}^{\infty }K_{dz}^{2}\left(
\tau \right) d\tau, \label{eq7}
\end{equation}
where
\begin{align}
K_{dz}\left(\tau \right)&= \left\{
\begin{array}{lr}
\frac{\sqrt{P}\kappa}{\sqrt{N}\lambda}\frac{1}{1+\sqrt{1+\Lambda }}
e^{-\lambda \left(\tau +\varepsilon\right)}&,\tau +\varepsilon\geq 0 \\
\frac{\sqrt{P}\kappa}{\sqrt{N}\lambda}\frac{1}{1+\sqrt{1+\Lambda }}
e^{\lambda\sqrt{1+\Lambda}\left(\tau+\varepsilon\right)}&,\tau
+\varepsilon <0 %
\end{array}.%
\right.
\end{align}
\begin{figure}[tbp]
\begin{center}
\includegraphics[width=80mm]{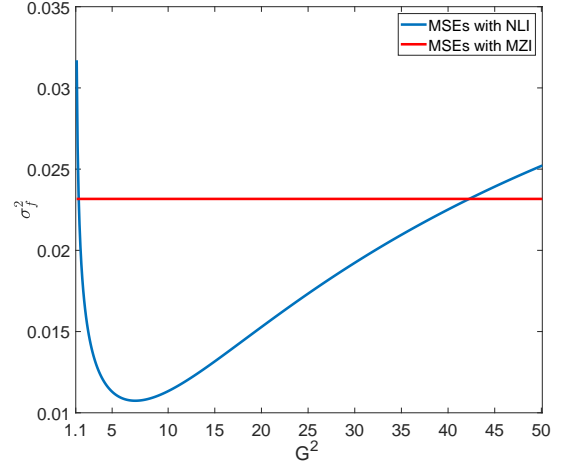}
\end{center}
\caption{The mean square error $\sigma _{f}^{2}$ of tracking as a function of $G^{2}$ for MZI (red line) and NLI (blue line). Parameters are $\kappa
=1.0\times 10^{4}rad/s, \lambda =1.0\times 10^{5}rad/s$, $\left\vert \beta \right\vert ^{2}=1.0\times 10^{7}s^{-1}$.}  \label{fig3}
\end{figure}
The stochastic phase is stationary and the optimum filter is time-invariant, so the minimum MSE is a constant.
To calculate the MSE analytically, we divide the discussion into three cases according to the value of $\varepsilon$: (i) $\varepsilon =0$, (ii) $\varepsilon >0$, (iii) $\varepsilon <0$. When $\varepsilon =0$, $ d\left( t\right) =\varphi \left( t\right) $, this is the phase tracking
case and $\xi=\sigma _{f}^{2}$. The integral result $\xi$ of Eq.(\ref{eq7}) is still implicit because $\Lambda $ is a function of $\sigma _{f}^{2}$. After solving the implicit result, the
minimum MSE of tracking is \cite{supplementary}
\begin{equation}
\sigma _{f}^{2}=\frac{-\left( \lambda -G^{2}g^{2}\kappa \right) +%
\sqrt{\left( \lambda -G^{2}g^{2}\kappa \right) ^{2}+4G^{2}g^{2}\left(
\frac{\left\vert \beta \right\vert ^{2}}{G^{2}+g^{2}}+\lambda \right) \kappa
} }{4G^{2}g^{2}\left( \frac{\left\vert \beta \right\vert ^{2}}{%
G^{2}+g^{2}}+\lambda \right) }. \label{eq9}
\end{equation}%
Similarly, the MSEs of the other two cases can be calculated as \cite{supplementary}
\begin{equation}
\xi_{NLI} =\left\{
\begin{array}{c}
\frac{\kappa }{2\lambda }\left[ 1-\frac{\Lambda }{\left( 1+\sqrt{
1+\Lambda }\right) ^{2}}e^{-2\lambda \varepsilon }\right] ,\varepsilon
> 0 \\
\frac{\kappa }{2\lambda }\left[\frac{1}{\sqrt{1+\Lambda }}+\frac{%
\Lambda e^{2\lambda \sqrt{1+\Lambda }\varepsilon }}{\left( 1+\sqrt{1+\Lambda
}\right) ^{2}\sqrt{1+\Lambda }}\right], \varepsilon
< 0 %
\end{array}%
\right.   ,\label{eq10}
\end{equation}
where $\varepsilon >0$ stands for predicting the future phase with current measurement outcomes and $\varepsilon <0$ is the case of smoothing.
Before investigating the enhancement of phase estimation with NLI, we set the precision of Ornstein-Uhlenback stochastic phase estimation with coherent state and MZI as the classical limit. To compare the MSE of the two types of interferometer, we make photon number flux inside them equivalent, i.e. $N_{MZI}=N_{NLI}=\left\vert \beta \right\vert ^{2}$. In this case the MSE of phase estimation with MZI can be written as \cite{supplementary}
\begin{equation}
\xi_{MZI} =\left\{
\begin{array}{c}
\frac{\kappa }{2\lambda }\left[ 1-\frac{\Lambda _{1}}{\left( 1+\sqrt{%
1+\Lambda _{1}}\right) ^{2}}e^{-2\lambda \varepsilon }\right] ,\varepsilon
> 0 \\
\frac{\kappa }{2\lambda }\{\frac{1}{\sqrt{1+\Lambda _{1}}}+\frac{\Lambda
_{1}e^{2\lambda \sqrt{1+\Lambda _{1}}\varepsilon }}{\left( 1+\sqrt{1+\Lambda
_{1}}\right) ^{2}\sqrt{1+\Lambda _{1}}}\},\varepsilon \leq 0%
\end{array}%
\right.   ,\label{eq11}
\end{equation}
where $\Lambda _{1}=\left\vert \beta \right\vert ^{2}\kappa/ \lambda ^{2}$.
To investigate the effect of the gain $G$ on the MSE $\sigma _{f}^{2},$ we consider the
stochastic phase tracking with a fixed photon number flux $\left\vert \beta \right\vert ^{2}=1.0\times
10^{7}s^{-1},$  $\kappa
=1.0\times 10^{4}rad/s, \lambda =1.0\times 10^{5}rad/s$ and vary $G^{2}$ from 1.1 to 50. Fig.\ref{fig3} shows that
the MSE $\sigma _{f}^{2}$  can be reduced by using the NLI compared to the classical limit with MZI. There is an optimal $G^{2}=7.4$ that gives the minimum MSE, which is feasible with current techniques \cite{Liu2018}. This is expected from the analysis of the SNR. Here, the optimal degree of gain depends on the  photon number flux $\left\vert \beta \right\vert ^{2}$, $\kappa$, and $\lambda$.
\begin{figure}[tbp]
\begin{center}
\includegraphics[width=80mm]{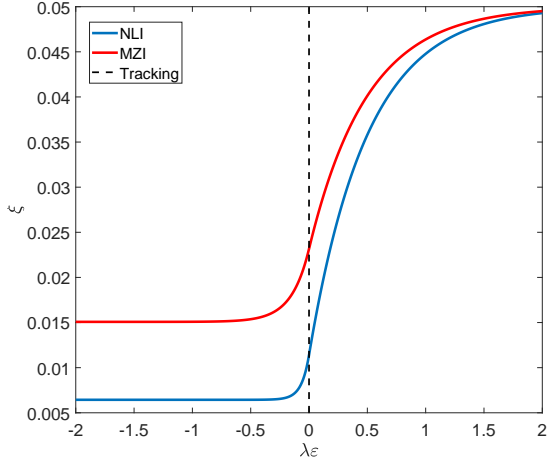}
\end{center}
\caption{The mean square error $\xi$ as a function of $\lambda \varepsilon$ for MZI (red line) and NLI (blue line). The horizontal axis is the proportion between $\varepsilon$ and the correlation time of $\varphi \left( t\right) $. The proportion equal to 0 represent phase tracking (black dotted line). $\lambda \varepsilon >0$ and $\lambda \varepsilon <0$ stand for  prediction and  smoothing respectively. Parameters are $\kappa
=1.0\times 10^{4}rad/s, \lambda =1.0\times 10^{5}rad/s$, $G^{2}=7.4$, $\left\vert \beta \right\vert ^{2}=1.0\times 10^{7}s^{-1}$.  }\label{fig4}
\end{figure}

Fig.\ref{fig4} shows the MSE $\xi$ as a function of $\lambda \varepsilon$ according to Eqs.(\ref{eq9})-(\ref{eq11}), where the horizontal axis is the ratio between $\varepsilon$ and the correlation time of $\varphi \left( t\right) $. Here we have set the parameters $\kappa
=1.0\times 10^{4}rad/s, \lambda =1.0\times 10^{5}rad/s$, $G^{2}=7.4$, $\left\vert \beta \right\vert ^{2}=1.0\times 10^{7}s^{-1}$. We can conclude three key implications from Fig.\ref{fig4}. Firstly, the MSE is increasing with the increase of $ \varepsilon$, and the smallest error is achieved with the smoothing. When $\varepsilon$ is close to the correlation time, the MSEs tend to be mean square variation of the stochastic phase $\kappa/2\lambda$, i.e., we can not predict the phase away from one coherent time. Secondly, the phase estimation with smoothing is nearly two times of tracking for both kinds of interferometers. Thirdly, the MSEs of all cases are reduced significantly below the classical limit (red line in Fig.\ref{fig4}) when we use the NLI.

\begin{figure}[h!]
\begin{center}
\includegraphics[width=80mm]{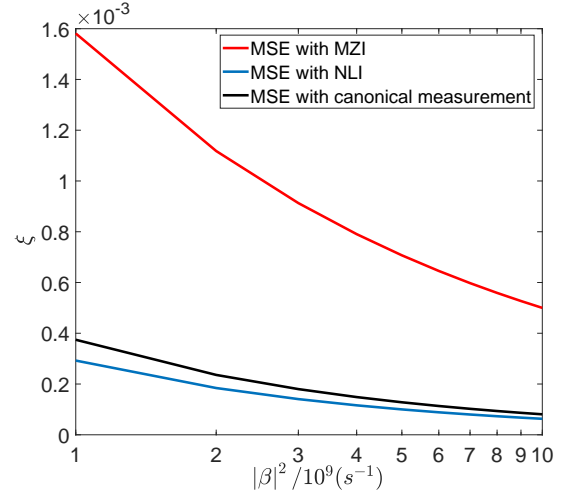}
\end{center}
\caption{ The optimal smoothing mean-square error $\xi$ as a function of
photon number flux $\left\vert \beta \right\vert ^{2}$ for MZI (Red line), NLI (Blue line) and  canonical measurement (Black line). Parameters are $\kappa
=1.0\times 10^{4}rad/s, \lambda =1.0\times 10^{5}rad/s$.} \label{fig5}
\end{figure}

So far we have shown that the stochastic phase estimation can be enhanced with NLI. In the following, we will demonstrate that the NLI achieves the Heisenberg scaling asymptotically.
For a fixed photon flux $\left\vert \beta \right\vert ^{2}$, the SNR of the measurement photocurrent Eq.(\ref{eq2}) is $SNR_{NLI}=\frac{4G^{2}\left( G^{2}-1\right) \left\vert \beta \right\vert
^{2}}{\left[ \left( 2G^{4}-G^{2}\right) \sigma _{f}^{2}+1\right] \left(
2G^{2}-1\right) }$. There is an optimal degree gain
$G_{o}$ maximizing the SNR and the minimum MSE of
phase tracking can be calculated as $\sigma _{f}^{2}\approx 1/{%
2G_{o}^{4}}$. When $G_{o}^{2}\gg 1,\frac{\left \vert \beta \right \vert ^{2}}{2G_{o}^{2}}\gg
\lambda ,\left \vert \beta \right \vert ^{2}\gg \kappa,G_{o}^{4}\kappa\gg \lambda$, the optimal gain $G_{o}$ meet the relationship $G_{o}^{2}\approx{
\left( \left\vert \beta \right\vert ^{2}\kappa ^{2}\right) ^{1/3}}/{%
2^{2/3}\kappa }$ and we can obtain the tracking MSE~\cite{supplementary}
\begin{equation}
\sigma _{f}^{2}\approx 2^{1/3}\left( \frac{\kappa }{\left\vert \beta
\right\vert ^{2}}\right) ^{2/3}.  \label{7}
\end{equation}
 Substituting this expression into Eq.(\ref{eq10}) yields the MSE of smoothing
\begin{equation}
\xi \approx \left( \frac{\kappa }{2\left\vert \beta \right\vert ^{2}}\right)
^{2/3},
\end{equation}
which means the MSE of stochastic phase estimation with NLI can achieve the stochastic HL scaling \cite{Berry2015,Berry2013,Dinani2017}. Fig.\ref{fig5} shows the optimal smoothing MSE in the two kinds of interferometers and canonical measurement for different mean photon flux, which varies from $\left\vert \beta \right\vert ^{2}= 10^{9}s^{-1}$ to $\left\vert \beta \right\vert ^{2}=10^{10}s^{-1}$. It can be seen that the phase estimation with NLI has an enhancement on scaling compared with the classical limit $\xi \approx \frac{1}{2}\left( \frac{\kappa }{\left\vert \beta \right\vert ^{2}}\right)^{1/2}$ using MZI, and the smoothing MSE can reach the stochastic Heisenberg scalling:
$O\left(\left( \frac{\kappa }{\left\vert \beta \right\vert ^{2}}\right)^{2/3}\right)$ \cite{Berry2013}. Moreover, using the NLI we can surpass the minimum MSE of canonical measurement which is $\frac{4}{5}\left( \frac{\kappa }{\left\vert \beta \right\vert ^{2}}\right)
^{2/3}$ \cite{Berry2015}.

\textit{Conclusion}$-$In summary, we have proposed the stochastic optical phase estimation with SU(1,1) interferometer. We find that a suitable range parametric amplification gain can enhance the estimation and there is an optimal gain minimizing the MSE. Moreover, compared with the classical limit with MZI, the mean square errors have significant reduction for prediction, tracking and smoothing simultaneously under the same photon number flux inside the interferometers if we optimize the parametric amplifier gain. At last, we can achieve the stochastic Heisenberg scaling, and surpass the minimum MSE using the canonical measurement. These results highlight the advantages of the SU(1,1) interferometer in stochastic optical phase estimation, and provide a new avenue for the practical quantum metrology.

\textbf{Acknowledgments} This work was supported by the National Key Research and Development Program of China (Grant Nos.2017YFA0303703 and 2019YFA0308704) and the National Natural Science Foundation of China (Grant Nos. 91836303, 61975077, 61490711, 11690032,11664017, 11874155 and 91436211), the Natural Science Foundation of Shanghai (Grant Nos. 17ZR1442900), the Nanjing University Innovation and Creative Program for PhD candidate (2016017).\newline

\begin{widetext}
\maketitle
\subsection{\label{sec:level6}The caculation of photocurrent}

The complete input-output relation of the NLI is
\begin{align}
\hat{c}_{out}=&G\left( G\hat{c}_{in}+g\hat{d}_{in}^{\dag }\right)
e^{i\Phi \left( t\right) }+g\left( g\hat{c}_{in}+G\hat{d}_{in}^{\dag
}\right) e^{-i\varphi \left( t\right) },  \nonumber \\
\hat{d}_{out}=&g\left( G\hat{c}_{in}^{\dag }+g\hat{d}_{in}\right)
e^{-i\Phi \left( t\right) }+G\left( g\hat{c}_{in}^{\dag }+G\hat{d}%
_{in}\right) e^{i\varphi \left( t\right) }.
\end{align}
When we perform homodyne detection at the output $d_{out}$, the measurement operator can be described as
\begin{equation}
\hat{X}_{dout}\left( \theta \left( t\right) \right) =\hat{d}%
_{out}^{\dagger }e^{i\theta \left( t\right) }+\hat{d}_{out}e^{-i\theta
\left( t\right) },
\end{equation}
where $\theta \left( t\right) $ is the phase of the local oscillator. If two input modes of the first PA are injected with a coherent state $\left\vert \alpha \right\rangle $ and
a vacuum state, the mean value of homodyne measurement is
\begin{align}
\left \langle \hat{X}_{dout}(\theta \left( t\right) )\right \rangle
&=\left \langle \hat{d}_{out}^{\dagger }e^{i\theta \left( t\right) }+%
\hat{d}_{out}e^{-i\theta \left( t\right) }\right \rangle  \nonumber \\
&= 4Gg\cos \frac{\left( \Phi \left( t\right) +\varphi \left( t\right)
\right) }{2}\cos \left( \frac{\Phi \left( t\right) -\varphi \left( t\right)
}{2}+\theta \left( t\right) \right) \left \vert \alpha \right \vert  \nonumber \\
&\approx 4Gg\sin \frac{\left( \varphi \left( t\right) -\varphi _{f}\left(
t\right) \right) }{2}\left \vert \alpha \right \vert  \nonumber \\
&\approx 2Gg\left( \varphi \left( t\right) -\varphi _{f}\left( t\right)
\right) \left \vert \alpha \right \vert,
\end{align}
and the variance is
\begin{align}
\Delta ^{2}\hat{X}_{dout}(\theta \left( t\right) ) &=\left \langle
\hat{X}_{dout}^{2}(\theta \left( t\right) )\right \rangle -\left \langle
\hat{X}_{dout}(\theta \left( t\right) )\right \rangle ^{2} \nonumber \\
&=4G^{2}g^{2}\left( 1+\cos \left( \Phi \left( t\right) +\varphi \left(
t\right) \right) \right) +1 \nonumber \\
&\approx 8G^{2}g^{2}\left( \sin ^{2}\frac{\left( \varphi \left( t\right) -\varphi
_{f}\left( t\right) \right) }{2}\right) +1 \nonumber \\
&\approx 2G^{2}g^{2}\left( \varphi \left( t\right) -\varphi _{f}\left(
t\right) \right) ^{2}+1.
\end{align}
The photocurrent that we interest in is
\begin{align}
X_{dout}\left( t\right)  &=\left \langle \hat{X}_{dout}(\theta \left(
t\right) )\right \rangle +\Delta \hat{X}_{dout}(\theta \left( t\right)
)n\left( t\right)  \nonumber \\
&\approx 2Gg\left( \varphi \left( t\right) -\varphi _{f}\left( t\right)
\right) \left \vert \alpha \right \vert +\sqrt{2G^{2}g^{2}\sigma _{f}^{2}+1}%
n(t),
\end{align}
where we have adaptively controlled the feedback phase in other
arm as $\Phi \left( t\right) =-\varphi _{f}\left( t\right) -\pi$, and the
phase of the local oscillator was controlled as $\theta \left( t\right)
=\varphi _{f}\left( t\right)+\pi/2$. $n\left( t\right)$ is a Gaussian white-noise term. Here, the photocurrent has Gaussian stationary statistic and $\sigma _{f}^{2}=\left\langle \left( \varphi \left( t\right) -\varphi_{f}\left( t\right) \right) ^{2}\right\rangle _{ss}$ is stationary MSE. When the
photocurrent is added by $2Gg\varphi _{f}\left( t\right)\left \vert \alpha \right \vert$, the
photocurrent can be approximately represented as
\begin{align}
r\left( t\right) &\approx 2Gg\left\vert \alpha \right\vert \varphi \left( t\right) +%
\sqrt{2G^{2}g^{2}\sigma _{f}^{2}+1}n\left( t\right)   \nonumber \\
&=\frac{2Gg\left\vert \beta \right\vert}{\sqrt{G^2+g^2}} \varphi \left( t\right) +%
\sqrt{2G^{2}g^{2}\sigma _{f}^{2}+1}n\left( t\right),
\end{align}
where we have defined that the  photon flux inside the interferometer is $\left\vert \beta \right\vert ^{2}=\left( G^{2}+g^{2}\right) \left\vert\alpha \right\vert ^{2}$.

In the MZI case, the splitting and recombining of light are accomplished by 50:50 beam spliters (BS). The relation of input-output of the beam spliter is $\hat{C}=1/\sqrt{2}\left(\hat{c}_{in}+i\hat{d}_{in}\right)$ and $\hat{D}=1/\sqrt{2}\left(i\hat{c}_{in}+\hat{d}_{in}\right)$. The complete input-output relation of the MZI is
\begin{align}
\hat{c}_{out}=\frac{1}{2}\left[ \left( e^{i\varphi \left( t\right) }-e^{i\Phi
\left( t\right) }\right) \hat{c}_{in}+i\left( e^{i\varphi \left( t\right) }+e^{i\Phi
\left( t\right) }\right) \hat{d}_{in}\right],  \nonumber \\
\hat{d}_{out}=\frac{1}{2}\left[ i\left( e^{i\varphi \left( t\right) }+e^{i\Phi
\left( t\right) }\right) \hat{c}_{in}-\left( e^{i\varphi \left( t\right) }-e^{i\Phi
\left( t\right) }\right) \hat{d}_{in}\right].
\end{align}
Similar to the NLI, two input modes of the first BS are injected with a coherent state $\left\vert \beta\right\rangle $ and
a vacuum state, the mean value of homodyne measurement is
\begin{align}
\left \langle \hat{X}_{dout}(\theta \left( t\right) )\right \rangle  &=\left \langle
\hat{d}_{out}^{\dagger }e^{i\theta \left( t\right) }+\hat{d}%
_{out}e^{-i\theta \left( t\right) }\right \rangle  \nonumber \\
&=\left[ 2\cos \frac{\left( \Phi \left( t\right) -\varphi \left( t\right)
\right) }{2}\cos \left( \frac{\Phi \left( t\right) +\varphi \left( t\right)
}{2}-\theta \left( t\right) +\frac{\pi }{2}\right) \right] \left \vert \beta
\right \vert  \nonumber \\
&\approx 2\left \vert \beta \right \vert \cos \left( \frac{\varphi
_{f}\left( t\right) +\varphi \left( t\right) }{2}-\theta \left( t\right) +%
\frac{\pi }{2}\right)  \nonumber \\
&\approx \left \vert \beta\right \vert \left( \varphi \left( t\right)
-\varphi _{f}(t)\right),
\end{align}
and the variance is
\begin{align}
\Delta ^{2}\hat{X}_{dout}(\theta \left( t\right) )=\left \langle \hat{%
X}_{dout}^{2}(\theta \left( t\right) )\right \rangle -\left \langle \hat{X}%
_{dout}(\theta \left( t\right) )\right \rangle ^{2}\approx 1.
\end{align}
The homodyne photocurrent is
\begin{align}
X_{dout}\left( t\right)  &=\left \langle \hat{X}_{dout}(\theta \left(
t\right) )\right \rangle +\Delta \hat{X}_{dout}(\theta \left( t\right)
)n\left( t\right)  \\
&\approx \left \vert \beta \right \vert \left( \varphi \left( t\right)
-\varphi _{f}\left( t\right) \right) +n(t).
\end{align}
Here, the two feedback phases is $\Phi \left( t\right) =\varphi _{f}\left( t\right) ;\theta \left( t\right)
=\varphi _{f}\left( t\right) +\pi$. When the
photocurrent is added by $\left \vert \beta \right \vert \varphi _{f}\left( t\right)  $, the
photocurrent can be approximately represented as
\begin{equation}
r\left( t\right) \approx \left \vert \beta \right \vert  \varphi \left( t\right)
  +n(t).
\end{equation}

\subsection{\label{sec:level2}Solution of Wiener-Hopf equation}

When we set $\tau =t-\sigma $ and $\upsilon =t-\epsilon ,$ the the Wiener-Hopf
equation
\begin{equation}
K_{dr}\left( t-\sigma \right) =\int_{-\infty }^{t}h_{o}\left( t-\epsilon
\right) K_{r}\left( \epsilon -\sigma \right) d\epsilon ,  \label{1}
\end{equation}%
became
\begin{equation}
K_{dr}\left( \tau \right) =\int_{0}^{\infty }h_{o}\left( \upsilon \right)
K_{r}\left( \tau -\upsilon \right) d\upsilon.
\end{equation}%
Here, we solve the equation with two steps, the first step of solving this equation is that suppose there is a whitening filter impulse
response $\omega \left( \tau,t \right) $ which can transfer $r\left( t\right) $
to white process $z\left( \tau\right) $ , the filtering process can be described
as
\begin{equation}
z\left( \tau \right) =\int_{-\infty }^{\infty }r\left( t\right) \omega
\left( \tau -t\right) dt.
\end{equation}%
 Taking the inverse Fourier transform on both sides we can obtain
\begin{equation}
\left\vert W\left( \omega \right) \right\vert ^{2}S_{r}\left( \omega \right)
=1,
\end{equation}%
where $W\left( \omega \right) $ is the transfer function of impulse response $%
\omega \left( \tau-t \right) $ and $S_{r}\left( \omega \right) $ is the
spectrum density of $r\left( t\right) .$ According to the equation (2) in the
main text we can calculate spectrum density as
\begin{equation}
S_{r}\left( \omega \right) =\frac{4G^{2}g^{2}\left\vert \alpha \right\vert
^{2}\kappa }{\omega ^{2}+\lambda ^{2}}+2G^{2}g^{2}\sigma _{f}^{2}+1,
\end{equation}
where we set $H^{+}\left( \omega \right) =\sqrt{N}\frac{i\omega +\lambda \sqrt{%
1+\Lambda }}{i\omega +\lambda },\Lambda =\frac{P\kappa }{N\lambda ^{2}}%
,N=2G^{2}g^{2}\sigma _{f}^{2}+1,P=4G^{2}g^{2}\left\vert \alpha \right\vert
^{2}$. The spectrum density can be decomposed as $S_{r}\left( \omega \right)=H^{+}\left( \omega \right)H\left( \omega \right)$. In this step, we can see that the transfer function which transfer $r\left( t\right) $
to white process $z\left( \tau\right) $ is $W\left( \omega \right)=\frac{1}{H^{+}\left( \omega \right) }$.

In the second step, we suppose $f_{o}\left( t,\tau \right) $ is impulse response of optimum linear filter for estimating
$d\left( t\right) $ with the $z\left( \tau\right) $, so the corresponding
Wiener-Hopf equation is
\begin{equation}
K_{dz}\left( \tau \right) =\int_{0}^{\infty }f_{o}\left( \upsilon \right)
K_{z}\left( \tau -\upsilon \right) d\upsilon ,\tau \geqslant 0.
\end{equation}%
Because $z\left( \tau\right) $ is white process, therefore $f_{o}\left( \tau
\right) =K_{dz}\left( \tau \right) $ and it can be found as
\begin{align}
K_{dz}\left( \tau \right) &= \left\langle d\left( t\right) \int_{-\infty
}^{\infty }\omega \left( \upsilon \right) r\left( t-\tau -\upsilon \right)
d\upsilon \right\rangle \nonumber \\
&=  \int_{-\infty }^{\infty }\omega \left( -\mu \right) K_{dr}\left( \tau
-\mu \right) d\mu,
\end{align}
if we take the inverse Fourier transform on both sides, we get
\begin{equation}
\left[ S_{dz}\left( \omega \right) \right] _{+}=\left[ W^{\ast }\left(
\omega \right) S_{dr}\left( \omega \right) \right] _{+}=\left[ \frac{%
S_{dr}\left( \omega \right) }{\left[ H^{+}\left( \omega \right) \right]
^{\ast }}\right] _{+},\label{eqs20}
\end{equation}%
where $\left[ S_{dz}\left( \omega \right) \right] _{+}=\int_{0}^{\infty
}K_{dz}\left( \tau \right) e^{-j\omega \tau }d\tau $. Here we use the subscript $[]_+ $ denotes that the integration time of the inverse transform is from 0 to $\infty $ . In this step, we can see that the transfer function of optimum linear filter for estimating $d\left( t\right) $ with the $z\left( \tau\right) $ is
\begin{equation}
F\left( \omega \right) =\left[ \frac{S_{dr}\left( \omega \right) }{\left[
H^{+}\left( \omega \right) \right] ^{\ast }}\right] _{+}.
\end{equation}%
After the two step, we can see the complete optimum linear processor in the frequency domain is
\begin{align}
H_{o}\left( \omega \right)= \frac{F(\omega )}{H^{+}\left( \omega \right) } =\frac{1}{H^{+}\left( \omega \right) }\left[
\frac{S_{dr}\left( \omega \right) }{\left[ H^{+}\left( \omega \right) %
\right] ^{\ast }}\right] _{+}.\label{eqs22}
\end{align}
On the other hand, there is correlation $K_{dr}\left( \tau \right)
=\left\langle d\left( t\right) r\left( t-\tau \right) \right\rangle =\sqrt{P}%
K_{\varphi }\left( \tau +\varepsilon \right) $ and $S_{dr}\left( \omega
\right) =\frac{\kappa \sqrt{P}e^{i\omega \varepsilon }}{\omega ^{2}+\lambda
^{2}}.$ So%
\begin{align}
S_{dz}\left( \omega \right)  &=\frac{S_{dr}\left( \omega \right) }{\left[
H^{+}\left( \omega \right) \right] ^{\ast }}=\frac{\kappa \sqrt{P}e^{i\omega
\varepsilon }}{\omega ^{2}+\lambda ^{2}}\frac{\lambda -i\omega }{\sqrt{N}%
\left( \lambda \sqrt{1+\Lambda }-i\omega \right) } \nonumber \\
&=\frac{\kappa \sqrt{P}e^{i\omega \varepsilon }}{\lambda +i\omega }\frac{1}{%
\sqrt{N}\left( \lambda \sqrt{1+\Lambda }-i\omega \right) }\nonumber \\
&=\frac{\kappa \sqrt{P}e^{i\omega \varepsilon }}{\sqrt{N}\lambda \left( 1+%
\sqrt{1+\Lambda }\right) }\left[ \frac{1}{\lambda +i\omega }+\frac{1}{\left(
\lambda \sqrt{1+\Lambda }-i\omega \right) }\right].
\end{align}
To find the $\left[ S_{dz}\left( \omega \right) \right] _{+}$, we take the
inverse Fourier transform
\begin{align}
K_{dz}\left( \tau \right)  &=\mathcal{F}^{-1}\left[ S_{dz}\left( \omega \right) %
\right] \nonumber \\
&=\mathcal{F}^{-1}\left[ \frac{\kappa \sqrt{P}e^{i\omega \varepsilon }}{\sqrt{N}%
\lambda \left( 1+\sqrt{1+\Lambda }\right) }\left[ \frac{1}{\lambda +i\omega }%
+\frac{1}{\left( \lambda \sqrt{1+\Lambda }-i\omega \right) }\right] \right]
\nonumber \\
&=\frac{\kappa \sqrt{P}e^{-\lambda \left( \tau +\varepsilon \right) }}{%
\sqrt{N}\lambda \left( 1+\sqrt{1+\Lambda }\right) }u\left( \tau +\varepsilon
\right) +\frac{\kappa \sqrt{P}e^{\lambda \sqrt{1+\Lambda }\left( \tau
+\varepsilon \right) }}{\sqrt{N}\lambda \left( 1+\sqrt{1+\Lambda }\right) }%
u\left( -\tau -\varepsilon \right), \label{eqs24}
\end{align}
where $u\left( \tau\right)$ is Heaviside function.
When $\varepsilon =0$,
\begin{equation}
f_{o}(\tau )=K_{dz}\left( \tau \right) =\frac{\kappa \sqrt{P}e^{-\lambda \tau }}{\sqrt{N}\lambda \left(
1+\sqrt{1+\Lambda }\right) }u\left( \tau \right),
\end{equation}
and
\begin{equation}
F(\omega )=\left[ S_{dz}\left( \omega \right) \right] _{+}=\frac{\kappa
\sqrt{P}}{\sqrt{N}\lambda \left( 1+\sqrt{1+\Lambda }\right) }\frac{1}{%
\lambda +i\omega },
\end{equation}
so the complete optimum linear processor of phase tracking in the frequency domain is
\begin{align}
H_{ot}(\omega )&=\frac{F(\omega )}{H^{+}\left( \omega \right) }
=\frac{%
\kappa \sqrt{P}}{\sqrt{N}\lambda \left( 1+\sqrt{1+\Lambda }\right) }\frac{1}{%
\lambda +i\omega }\frac{\lambda +i\omega }{\sqrt{N}\left( \lambda \sqrt{%
1+\Lambda }+i\omega \right) } \nonumber \\
&=\frac{\kappa \sqrt{P}}{N\lambda \left( 1+\sqrt{1+\Lambda }\right) \left(
\lambda \sqrt{1+\Lambda }+i\omega \right) }.
\end{align}
When $\varepsilon <0$,
\begin{equation}
F(\omega )=\left[ S_{dz}\left( \omega \right) \right] _{+}=\frac{\kappa
\sqrt{P}}{\sqrt{N}}\left[ \frac{e^{i\omega \varepsilon }}{\left( \lambda
+i\omega \right) \left( \lambda \sqrt{1+\Lambda }-i\omega \right) }-\frac{%
e^{\varepsilon \lambda \sqrt{1+\Lambda }}}{\lambda \left( 1+\sqrt{1+\Lambda }%
\right) \left( \lambda \sqrt{1+\Lambda }-i\omega \right) }\right].
\end{equation}
So the complete optimum linear processor of smoothing in the frequency domain is
\begin{align}
H_{os}(\omega )&=\frac{F(\omega )}{H^{+}\left( \omega \right) }
=\frac{\kappa \sqrt{P}}{\sqrt{N}}\left[ \frac{e^{i\omega \varepsilon }}{%
\left( \lambda +i\omega \right) \left( \lambda \sqrt{1+\Lambda }-i\omega
\right) }-\frac{e^{\varepsilon \lambda \sqrt{1+\Lambda }}}{\lambda \left( 1+%
\sqrt{1+\Lambda }\right) \left( \lambda \sqrt{1+\Lambda }-i\omega \right) }%
\right] \frac{\lambda +i\omega }{\sqrt{N}\left( \lambda \sqrt{1+\Lambda }%
+i\omega \right) }  \nonumber \\
&=\frac{\kappa \sqrt{P}e^{i\omega \varepsilon }}{N\left( \lambda ^{2}\left(
1+\Lambda \right) +\omega ^{2}\right) }\left[ 1-\frac{e^{\varepsilon \left(
\lambda \sqrt{1+\Lambda }-i\omega \right) }\left( \lambda +i\omega \right) }{%
\lambda \left( 1+\sqrt{1+\Lambda }\right) }\right].
\end{align}
When $\varepsilon >0$,
\begin{equation}
F(\omega )=\left[ S_{dz}\left( \omega \right) \right] _{+}=\frac{\kappa
\sqrt{P}e^{^{i\omega \varepsilon }}}{\sqrt{N}\lambda \left( 1+\sqrt{%
1+\Lambda }\right) }\frac{1}{\lambda +i\omega },
\end{equation}
so the complete optimum linear processor of prediction in the frequency domain is
\begin{align}
H_{op}(\omega ) &=\frac{F(\omega )}{H^{+}\left( \omega \right) }=\frac{%
\kappa \sqrt{P}e^{^{i\omega \varepsilon }}}{\sqrt{N}\lambda \left( 1+\sqrt{%
1+\Lambda }\right) }\frac{1}{\lambda +i\omega }\frac{\lambda +i\omega }{%
\sqrt{N}\left( \lambda \sqrt{1+\Lambda }+i\omega \right) }  \nonumber \\
&=\frac{\kappa \sqrt{P}e^{^{i\omega \varepsilon }}}{N\lambda \left( 1+\sqrt{%
1+\Lambda }\right) \left( \lambda \sqrt{1+\Lambda }+i\omega \right) }.
\end{align}
Similarly, for MZI case, the complete optimum linear processor in the frequency domain is
\begin{align}
H_{o}(\omega )&= \left\{
\begin{array}{lr}
\frac{\kappa\left\vert \beta \right\vert e^{^{i\omega\varepsilon}}}{\lambda\left(1+\sqrt{
1+\Lambda_{1}}\right)\left(\lambda\sqrt{1+\Lambda_{1}}+i\omega\right)}&,\varepsilon >0 \\%
\frac{\kappa\left\vert \beta \right\vert e^{i\omega\varepsilon}}{\left(\lambda^{2}\left(
1+\Lambda_{1} \right)+\omega^{2}\right)}\left[1-\frac{e^{\varepsilon\left(
\lambda\sqrt{1+\Lambda_{1}}-i\omega\right)}\left(\lambda+i\omega\right)}{
\lambda\left(1+\sqrt{1+\Lambda_{1}}\right)}\right]&,\varepsilon \leq 0
\end{array},
\right.   \label{eq6}
\end{align}
where we set $\Lambda_{1} =\frac{\left\vert \beta \right\vert^2 \kappa }{\lambda ^{2}}.$

\subsection{\label{sec:level2}The calculation of minimum mean square error}

In this paper, the phase to be estimated is stationary and the optimum
filter is time-invariant, so the MSE is time independent and can be calculated as
\begin{align}
\xi \left( t\right) &=\left\langle[ d(t)-\int_{-\infty }^{t}r\left( \tau
\right) h_{o}\left( t-\tau \right) d\tau]^2 \right\rangle \nonumber \\
&=K_{d}\left( 0\right) -\int_{-\infty }^{t}h_{o}\left( t-\tau \right)
K_{dr}\left( t-\tau \right) d\tau \nonumber \\
&=K_{d}\left( 0\right) -\int_{0}^{\infty }h_{o}\left( \gamma \right)
K_{dr}\left( \gamma \right) d \gamma  \nonumber \\
&=K_{d}\left( 0\right) -\int_{0}^{\infty }K_{dz}\left( t\right) dt\left[
\frac{1}{2\pi }\int_{-\infty }^{\infty }e^{-j\omega t}d\omega \frac{1}{%
H^{+}\left( \omega \right) }\int_{-\infty }^{\infty }K_{dr}\left( \tau
\right) e^{j\omega \tau }d\tau \right],
\end{align}
in third line we let $t-\tau=\gamma $, and we substitue $h_{o}\left(
\gamma \right) $ with the inverse transform of $H_{o}\left( \omega
\right) =\frac{1}{H^{+}\left( \omega \right) }\int_{0}^{\infty }K_{dz}\left(
t\right) e^{-j\omega t}dt$ in last line. Moreover, from Eq. (\ref{eqs20}) we can see $K_{dz}\left(
t\right) =\frac{1}{2\pi }\int_{-\infty }^{\infty }e^{-j\omega t}d\omega
\frac{1}{H^{+}\left( \omega \right) }\int_{-\infty }^{\infty }K_{dr}\left(
\tau \right) e^{j\omega \tau }d\tau $.
So the MSE can be expressed as
\begin{equation}
\xi \left( t\right) =K_{d}\left( 0\right) -\int_{0}^{\infty
}K_{dz}^{2}\left( t\right) dt,
\end{equation}%
where $K_{d}\left( 0\right) =\frac{\kappa }{2\lambda }$ and $K_{dz}$ can be found from Eq.(\ref{eqs24}). When $\varepsilon =0,$ filtering with zero delay, which is the phase tracking case. The integral result is
\begin{align}
\sigma _{f}^{2} &=\frac{\kappa }{2\lambda }-\int_{0}^{\infty }\frac{P\kappa
^{2}}{N\lambda ^{2}}\frac{1}{\left( 1+\sqrt{1+\Lambda }\right) ^{2}}%
e^{-2\lambda \tau }d\tau \nonumber \\
&=\frac{\kappa }{2\lambda }\left[ 1-\frac{\Lambda }{\left( 1+\sqrt{%
1+\Lambda }\right) ^{2}}\right],
\end{align}
which still implicitly because $\Lambda $ is a function of $\sigma _{f}^{2}.$
After solving the implicitly result, the MSE of phase
tracking is
\begin{equation}
\sigma _{f}^{2}=\frac{\left[-\left( \lambda -G^{2}g^{2}\kappa \right) +%
\sqrt{\left( \lambda -G^{2}g^{2}\kappa \right) ^{2}+4G^{2}g^{2}\left(
\frac{\left\vert \beta \right\vert ^{2}}{G^{2}+g^{2}}+\lambda \right) \kappa
}\right] }{4G^{2}g^{2}\left( \frac{\left\vert \beta \right\vert ^{2}}{%
G^{2}+g^{2}}+\lambda \right) }.
\end{equation}%
When $\varepsilon >0,$ filterig with prediction, and the MSE of the prediction is
\begin{align}
\xi _{p} &=\frac{\kappa }{2\lambda }-\int_{0}^{\infty }\frac{P\kappa ^{2}}{%
N\lambda ^{2}}\frac{1}{\left( 1+\sqrt{1+\Lambda }\right) ^{2}}e^{-2\lambda
\left( \tau +\varepsilon \right) }d\tau \nonumber \\
&=\frac{\kappa }{2\lambda }\left[ 1-\frac{\Lambda }{\left( 1+\sqrt{%
1+\Lambda }\right) ^{2}}e^{-2\lambda \varepsilon }\right].
\end{align}
When $\varepsilon <0,$ it is the case smoothing. The MSE of
smoothing is
\begin{align}
\xi _{s} &=\frac{\kappa }{2\lambda }-\frac{P\kappa ^{2}}{N\lambda ^{2}}%
\frac{1}{\left( 1+\sqrt{1+\Lambda }\right) ^{2}}\left[ \int_{0}^{-%
\varepsilon }e^{2\lambda \sqrt{1+\Lambda }\left( \tau +\varepsilon \right)
}d\tau +\int_{-\varepsilon }^{\infty }e^{-2\lambda \left( \tau +\varepsilon
\right) }d\tau \right]   \nonumber \\
&=\frac{\kappa }{2\lambda }-\frac{P\kappa ^{2}}{N\lambda ^{2}}\frac{1}{%
\left( 1+\sqrt{1+\Lambda }\right) ^{2}}\left[ \frac{\left( 1-e^{2\lambda
\sqrt{1+\Lambda }\varepsilon }\right) }{2\lambda \sqrt{1+\Lambda }}+\frac{1}{%
2\lambda }\right]   \nonumber \\
&=\frac{\kappa }{2\lambda }\left[ \frac{1}{\sqrt{1+\Lambda }}+\frac{\Lambda
e^{2\lambda \sqrt{1+\Lambda }\varepsilon }}{\left( 1+\sqrt{1+\Lambda }%
\right) ^{2}\sqrt{1+\Lambda }}\right].
\end{align}
For the MZI case,  we set $H^{+}\left( \omega \right) =\sqrt{N}\frac{i\omega +\lambda \sqrt{
1+\Lambda_1 }}{i\omega +\lambda },\Lambda_1 =\frac{P\kappa }{N\lambda ^{2}},N=1,P= \left\vert \beta \right\vert^{2}$ and use the same calculation method with the MLI. The MSE of phase estimation with MZI can be written as
\begin{equation}
\xi_{MZI} =\left\{
\begin{array}{c}
\frac{\kappa }{2\lambda }\left[ 1-\frac{\Lambda _{1}}{\left( 1+\sqrt{%
1+\Lambda _{1}}\right) ^{2}}e^{-2\lambda \varepsilon }\right] ,\varepsilon
> 0 \\
\frac{\kappa }{2\lambda }\{\frac{1}{\sqrt{1+\Lambda _{1}}}+\frac{\Lambda
_{1}e^{2\lambda \sqrt{1+\Lambda _{1}}\varepsilon }}{\left( 1+\sqrt{1+\Lambda
_{1}}\right) ^{2}\sqrt{1+\Lambda _{1}}}\},\varepsilon \leq 0%
\end{array}%
\right.   \label{eqS}
\end{equation}

\subsection{\label{sec:level2}The stochastic Heisenberg limit with NLI}

The signal to noise of the photocurrent Eq.(2) in the mian text is

\begin{equation}
SNR_{NLI}=\frac{4G^{2}g^{2}\left\vert \alpha \right\vert ^{2}}{2G^{2}g^{2}\sigma
_{f}^{2}+1}=\frac{4G^{2}\left( G^{2}-1\right) \left\vert \beta \right\vert
^{2}}{\left( \left( 2G^{4}-G^{2}\right) \sigma _{f}^{2}+1\right) \left(
2G^{2}-1\right) }.
\end{equation}
The optimal gain $G_{o}$ that minimize the MSE of estimation equal to that
maximize the signal to noise. Taking the derivative of both sides of the
equation with respect to $G^{2},$ and let the derivative to be $0,$ we can
obtain
\begin{equation}
\sigma _{f}^{2}=\frac{4\left[ 2G_{o}^{2}\left( G_{o}^{2}-1\right) +1\right]
}{\left[ 4G_{o}^{2}\left( G_{o}^{2}-1\right) \right] ^{2}}\approx \frac{1}{2G_{o}^{4} }.
\end{equation}
The approximate equals sign is true in the case of $G_{o} ^{2}\gg 1$. Then it is combined with Eq.(9) in main text , we can obtain
\begin{align}
\frac{1}{2G_{o}^{4}} &=\frac{1}{8\left(
G_{o}^{4}-G_{o}^{2}\right) \left( \frac{\left\vert \beta \right\vert ^{2}}{%
2G_{o}^{2}-1}+\lambda \right) } \nonumber \\
&\left[ -2\left( \lambda -\left( G_{o}^{4}-G_{o}^{2}\right) \kappa \right) +%
\sqrt{4\left( \lambda -\left( G_{o}^{4}-G_{o}^{2}\right) \kappa \right)
^{2}+16\left( G_{o}^{4}-G_{o}^{2}\right) \left( \frac{\left\vert \beta
\right\vert ^{2}}{2G_{o}^{2}-1}+\lambda \right) \kappa }\right].
\end{align}
When $G_{o}^{2}\gg 1,\frac{\left \vert \beta \right \vert ^{2}}{2G_{o}^{2}}\gg
\lambda ,\left \vert \beta \right \vert ^{2}\gg \kappa,G_{o}^{4}\kappa\gg \lambda$, it can transform to
\begin{equation}
\frac{1}{2G_{o}^{4}}\approx \frac{\left[ -2\left( -\left( G_{o}^{4}\right)
\kappa \right) +\sqrt{4\left( -\left( G_{o}^{4}\right) \kappa \right)
^{2}+8G_{o}^{2}\left \vert \beta \right \vert ^{2}\kappa }\right] }{8\left(
G_{o}^{4}\right) \left( \frac{\left \vert \beta \right \vert ^{2}}{2G_{o}^{2}}%
\right) }.
\end{equation}
After the calculation, the optimal gain $G_{o}$ is
\begin{equation}
G_{o}\approx \sqrt{\frac{\left( \left\vert \beta \right\vert ^{2}\kappa
^{2}\right) ^{1/3}}{2^{2/3}\kappa }}.
\end{equation}
Substituting it to Eq.(S41), we can obtain the tracking MSE
\begin{equation}
\sigma _{f}^{2}\approx 2^{1/3}\left( \frac{\kappa }{\left\vert \beta
\right\vert ^{2}}\right) ^{2/3}.
\end{equation}
\end{widetext}
\end{document}